\documentstyle[12pt,cite,epsf]{article}
\renewcommand{\thefootnote}{\fnsymbol{footnote}}
\def \be{\begin{equation}}
\def \ee{\end{equation}}
\def \bea{\begin{eqnarray}}
\def \eea{\end{eqnarray}}
\def \bem#1{\renewcommand{\thefootnote}{\arabic{footnote}}\footnote{#1}}

\def \braket#1#2#3{\left\langle #1\right|#2\left| #3\right\rangle}

\def \eq#1{Eq.~(\ref{#1})}
\def \eqs#1#2{Eqs.~(\ref{#1})--(\ref{#2})}
\def \fig#1{Fig.~\ref{#1}}

\def \fqm{{\mathrm FQM}}
\def \ha{Hamiltonian}
\def \half{{1\over 2}}
\def \heff{H_{\mathrm{eff}}}
\def \H{heavy quark effective theory}
\def \HQ{HQET}
\def \Im{{\mathrm{Im}}\,}

\def \LB{\Lambda_b\to X_s\,\gamma}

\def \nnu{\nonumber}

\def \rf{Ref.~\cite}
\def \rfs{Refs.~\cite}
\def \sec#1{Sec.~\ref{#1}}
\def \SI{\sum \limits_{X_s,\,{\mathrm{pol}}}}

\def \Vec#1{\mbox{\bf#1}}
\def\today{\ifcase\month\or
  January\or February\or March\or April\or May\or June\or
  July\or August\or September\or October\or November\or December\fi
  \space\number\year}
\newcounter{Section}
\def \theSection{\Roman{Section}}
\newcommand{\Sec}[1]{\refstepcounter{Section}%
\centerline{\bf\theSection. #1}\setcounter{equation}{0}}

\newcounter{Subsection}[Section]

\newcounter{Subsubsection}[Section]

\newcounter{Appendix}

\def\ijmp#1#2#3{{\it Int.~J.~Mod.~Phys.\/}~{\bf A#1} (19#2) #3}

\def\np#1#2#3{{\it Nucl.~Phys.\/}~{\bf B#1} (19#2) #3}
\def\pl#1#2#3{{\it Phys.~Lett.\/}~{\bf B#1} (19#2) #3}
\def\pp{{\it preprint\/} }
\def\prd#1#2#3{{\it Phys.~Rev.\/}~{\bf D#1} (19#2) #3}

\def\prp#1#2#3{{\it Phys.~Rep.\/}~{\bf #1} (19#2) #3}

\begin{document}
\begin{flushright}
PITHA 95/12 \\ hep-ph/9505354 \\ May 1995
\end{flushright}
\vspace{0.5cm}
\begin{center}
\LARGE \bf Angular Distribution and Polarization of Photons in the
Inclusive Decay $\Lambda_b\to X_s\,\gamma$
\end{center}
\vspace{0.05cm}
\begin{center}\sc M.~Gremm\footnote{\footnotesize
Electronic address: gremm@physik.rwth-aachen.de},
F.~Kr\"uger\footnote{\footnotesize
Electronic address: krueger@physik.rwth-aachen.de} and
L.\,M.~Sehgal\footnote{\footnotesize Electronic address:
sehgal@physik.rwth-aachen.de}
\\ \it III. Physikalisches Institut (A), RWTH Aachen\\
D-52074 Aachen, Germany
\end{center}
\vspace{1.0cm}
\thispagestyle{empty}
\centerline{\bf ABSTRACT}
\begin{quotation}
We study the angular distribution of photons produced in the inclusive decay
of a polarized $\Lambda_b$, $\Lambda_b\to X_s\,\gamma$, using the technique of
\H.
Finite non-perturbative corrections are obtained relative to the free quark
decay $b\to s\gamma$. These corrections affect significantly the intensity
and polarization of photons emitted at small angles relative to the $\Lambda_b$
spin direction.
\end{quotation}
\setcounter{footnote}{0}
%
%
\newpage
\Sec{INTRODUCTION} \label{intro}
\bigskip
The decay $b\to s\gamma$, calculated on the level of free quarks, has three
interesting features:
\begin{enumerate}
\item[(i)] The photon is monochromatic, its energy spectrum being
\be\label{fqm1}
\left(\frac{1}{\Gamma}\frac{d\Gamma}{dy}\right)_\fqm = \delta(y-y_0)\ ,
\quad y=\frac{2E_{\gamma}}{m_b}\ , \quad y_0=\left(1-\frac{m_s^2}{m_b^2}
\right)\ .
\ee
\item[(ii)] The photon is emitted preferentially backwards relative to the spin
of the $b$-quark, the angular distribution being
\be\label{fqm2}
\left(\frac{1}{\Gamma}\frac{d\Gamma}{d\cos\theta}\right)_\fqm = \half
\left(1-\frac{m_b^2-m_s^2}{m_b^2+m_s^2}\cos\theta \right)\ .
\ee
\item[(iii)] The photon is predominantly left-handed, the angular distribution
for helicities $\lambda_{\gamma}=\pm1$ being
\bea\label{fqm3}
\left(\frac{1}{\Gamma}\frac{d\Gamma_+}{d\cos\theta}\right)_\fqm&=&
\half\,\frac{m_s^2}{m_b^2}\,(1+\cos\theta)
\left(1+\frac{m_s^2}{m_b^2}\right)^{-1}
\nnu \ ,\\
\left(\frac{1}{\Gamma}\frac{d\Gamma_-}{d\cos\theta}\right)_\fqm&=&
\half\,(1-\cos\theta)\left(1+\frac{m_s^2}{m_b^2}\right)^{-1}\ ,\\
P\equiv\frac{\Gamma_+-\Gamma_-}{\Gamma_++\Gamma_-}&=&-\frac{m_b^2-m_s^2}
{m_b^2+m_s^2}\nnu \ .
\eea
\end{enumerate}
Whereas feature (i) is purely kinematical (reflecting the two-body nature
of the decay), features (ii) and (iii) are specific consequences of the
standard model, in which the effective \ha\ governing $b\to s\gamma$ has the
structure
\bea\label{effham}
\heff&=& \frac{-4G_F}{\sqrt{2}}\frac{e}{16\pi^2}\,V_{tb}^{}V_{ts}^*\,c_7(m_b)\,
\bar{s}\,\sigma^{\mu\nu}\left(m_b P_R+m_s P_L\right)b F_{\mu\nu}\nnu \\
&\equiv&\frac{-4G_F}{\sqrt{2}}\frac{e}{16\pi^2}\,V_{tb}^{}V_{ts}^*\,c_7(m_b)\,
\bar{s}\,\Gamma^{\mu\nu}\,bF_{\mu\nu}\ ,
\eea
leading to a decay width
\be\label{decayfqm}
\Gamma_\fqm (b\to s\gamma)=\frac{\alpha G_F^2 m_b^5}{32 \pi^4}
\left|V_{tb}^{}V_{ts}^*\right|^2 \left|c_7(m_b)\right|^2
(1+\frac{ m_s^2}{m_b^2})(1-\frac{m_s^2}{m_b^2})^3\ .
\ee

The purpose of this paper is to study how the characteristics of
$b\to s\gamma$, summarized in \eqs{fqm1}{fqm3}, are altered when the $b$-quark
is embedded in a polarized $\Lambda_b$ baryon, and when the final $s$-quark
is a part of a hadronic system that is summed over. These changes have their
origin in the ``Fermi-motion'' of the $b$-quark within the hadron, as well
as its spin-dependent interaction with the environment, and can be parametrized
using the method of heavy quark effective theory (HQET) \cite{rev,cgg}.
In \sec{technique}, we outline the method of calculation, and discuss our
results in \sec{result}.
\vspace{2cm}

\Sec{\HQ\ AND THE DECAY $\Lambda_b\to X_s\,\gamma$}\label{technique}
\bigskip

The differential decay rate for the inclusive decay $\Lambda_b(p)\to X_s(p_X)
\, \gamma(p_{\gamma})$ in the standard model can be written as
\bea\label{rate}
d\Gamma&=&\SI(2\pi)^4\delta^4(p-p_X-p_{\gamma})\,\frac{d^3\Vec{p}_X}{(2\pi)^3
2E_X}\nnu \\
& & \times\braket{\Lambda_b}{\heff^{\dagger}(0)}{X_s \gamma}
\braket{X_s \gamma}{\heff(0)}{\Lambda_b}\frac{d^3\Vec{p}_{\gamma}}{(2\pi)^3
2E_{\gamma}}\ ,
\eea
where the effective \ha\ $\heff$ is given in (\ref{effham}). Using the optical
theorem, and writing the phase space element as
\be
\frac{{d^3\Vec{p}_{\gamma}}}{(2\pi)^3 2E_{\gamma}} =
\frac{m_b^2}{32\pi^2}\,y\,dy
\,d\cos\theta\ ,
\ee
\eq{rate} can be rewritten in the form
\be\label{rate1}
\frac{d\Gamma}{dy\,d\cos\theta}=\frac{\alpha G_F^2 m_b^2}{2^7 \pi^5}
\left|V_{tb}^{}V_{ts}^*\right|^2 \left|c_7(m_b)\right|^2 y\,\Im
T(y,\cos\theta)\ ,
\ee
where $T$ is given by
\bea
T=E_{\mu\nu\alpha\beta}T^{\mu\nu\alpha\beta}\nnu\ ,
\eea
with
\be
E_{\mu\nu\alpha\beta}\equiv \sum\limits_{\lambda}\braket{0}{F_{\mu\nu}}{\gamma}
\braket{\gamma}{F_{\alpha\beta}}{0}\ ,
\ee
and
\be\label{amplitude}
T_{\mu\nu\alpha\beta}= i\int d^4x\,\,e^{-ip_{\gamma}\cdot x}\braket{\Lambda_b}
{{\mathrm T}\left\{\bar{b}(x){{\Gamma^{\dagger}_{\mu\nu}}}s(x),
\bar{s}(0){\Gamma_{\alpha\beta}}b(0)\right\}}{\Lambda_b}\ .
\ee

The time-ordered product appearing in $T_{\mu\nu\alpha\beta}$ can be expanded
in
powers of $1/m_b$, using methods described in \cite{fls,fn,mw,mn,bsv,cgg,tm}.
To order $1/m_b^2$, we obtain
\bea\label{amplitude1}
&&T(y,\cos\theta)=2y^2m_b^3\,\frac{1}{(y-y_0-i\epsilon)}\nnu \\
&\times&\left\{\left[1+K\left(\frac{5}{3}
-\frac{7}{3}\,\frac{y}{(y-y_0-i\epsilon)}+\frac{2}{3}\,
\frac{y^2}{(y-y_0-i\epsilon)^2}
\right)\right]\left(1+\frac{m_s^2}{m_b^2}\right)\right.\nnu \\
&-&\left.\cos\theta\left[1+\epsilon_b+K\left(\frac{5}{3}
-\frac{7}{3}\,\frac{y}{(y-y_0-i\epsilon)}+\frac{2}{3}\,
\frac{y^2}{(y-y_0-i\epsilon)^2}
\right)\right]\left(1-\frac{m_s^2}{m_b^2}\right)\right\}\ .\nnu \\
\eea
The imaginary part $\Im\,T(y,\cos\theta)$ is then obtained by the formal
replacement
\bea\label{deltafunc}
&\displaystyle\frac{1}{{y-y_0-i\epsilon}}& \quad \to \quad \pi\,\delta(y-y_0)
\nnu\ ,\\
&\displaystyle\frac{1}{(y-y_0-i\epsilon)^2}& \quad \to \quad
-\pi\,\delta'(y-y_0)
\ ,\\
&\displaystyle\frac{1}{(y-y_0-i\epsilon)^3}& \quad \to \quad \frac{\pi}{2}
\,\delta''(y-y_0)\nnu\ .
\eea
The distributions in \eq{deltafunc} yield physically meaningful results only
if one integrates over the photon energy (see e.g. \rf{mn}).
The leading term proportional to $\delta(y-y_0)$ reproduces the result of
the free quark model (FQM). The corrections to the free quark result, involving
$\delta'(y-y_0)$ and $\delta''(y-y_0)$, are parametrized by two
phenomenological constants $K$ and $\epsilon_b$, defined as
\bea
K&=&-\braket{\Lambda_b}{\bar h_v\frac{(iD)^2}{2m_b^2}h_v}{\Lambda_b}\ ,\\
(1+\epsilon_b)s^{\mu}&=&\braket{\Lambda_b}{\bar b\,\gamma^{\mu}\gamma^5b}
{\Lambda_b}\ .
\eea
(\rfs{fls,mn} use a parameter $\lambda_1$ related to $K$ by
$K=-\lambda_1/{2m_b^2}$).
Integration of \eq{rate1} over $y$ then yields the angular distribution
\bea\label{decay}
\frac{d\Gamma(\LB)}{d\cos\theta}=\half\Gamma_\fqm
\left[(1-K) -(1+\epsilon_b -K)\,\frac{m_b^2-m_s^2}{m_b^2+m_s^2}\,\cos\theta
\right]\ .
\eea
If the decay $\LB$ is calculated for a fixed photon helicity
$(\lambda_{\gamma}=\pm 1)$, the angular distributions are
\be\label{hel+}
\frac{d\Gamma_+(\LB)}{d\cos\theta}=\half\Gamma_\fqm\left(\frac{m_s^2}{m_b^2}\right)\left[(1-K)+(1+\epsilon_b-K)\cos\theta\right]\left(1+\frac{m_s^2}{m_b^2}
\right)^{-1}\ ,
\ee
\be\label{hel-}
\frac{d\Gamma_-(\LB)}{d\cos\theta}=\half\Gamma_\fqm\left[(1-K)-(1+\epsilon_b-K)
\cos\theta\right]\left(1+\frac{m_s^2}{m_b^2}\right)^{-1}\ .
\ee
\eqs{decay}{hel-} are the HQET analogs of the free quark result
(\ref{fqm2})--(\ref{fqm3}).
\vspace{2cm}

\Sec{DISCUSSION OF RESULTS}\label{result}
\bigskip

To evaluate the HQET corrections, we need an estimate of the parameters $K$
and $\epsilon_b$. Following \rf{mw} we use $K\simeq0.01$, although this
parameter has a large uncertainty.
{}From the fact that $d\Gamma/d\cos\theta \ge 0$ for all scattering angles
(independent of the value of $m_s$ or $K$) we must have
$\epsilon_b <0$ (see \eq{decay}). We will use the value
$\epsilon_b=-\frac{2}{3}K$
suggested in \rf{fn}.\bem{This value is obtained if one neglects the
contributions coming from the double insertion of the chromomagnetic
operator \cite{fn}.} This is in agreement with the bound derived in \rf{kp}.

\begin{figure}[b]
\centerline{\epsfysize=12cm\epsffile{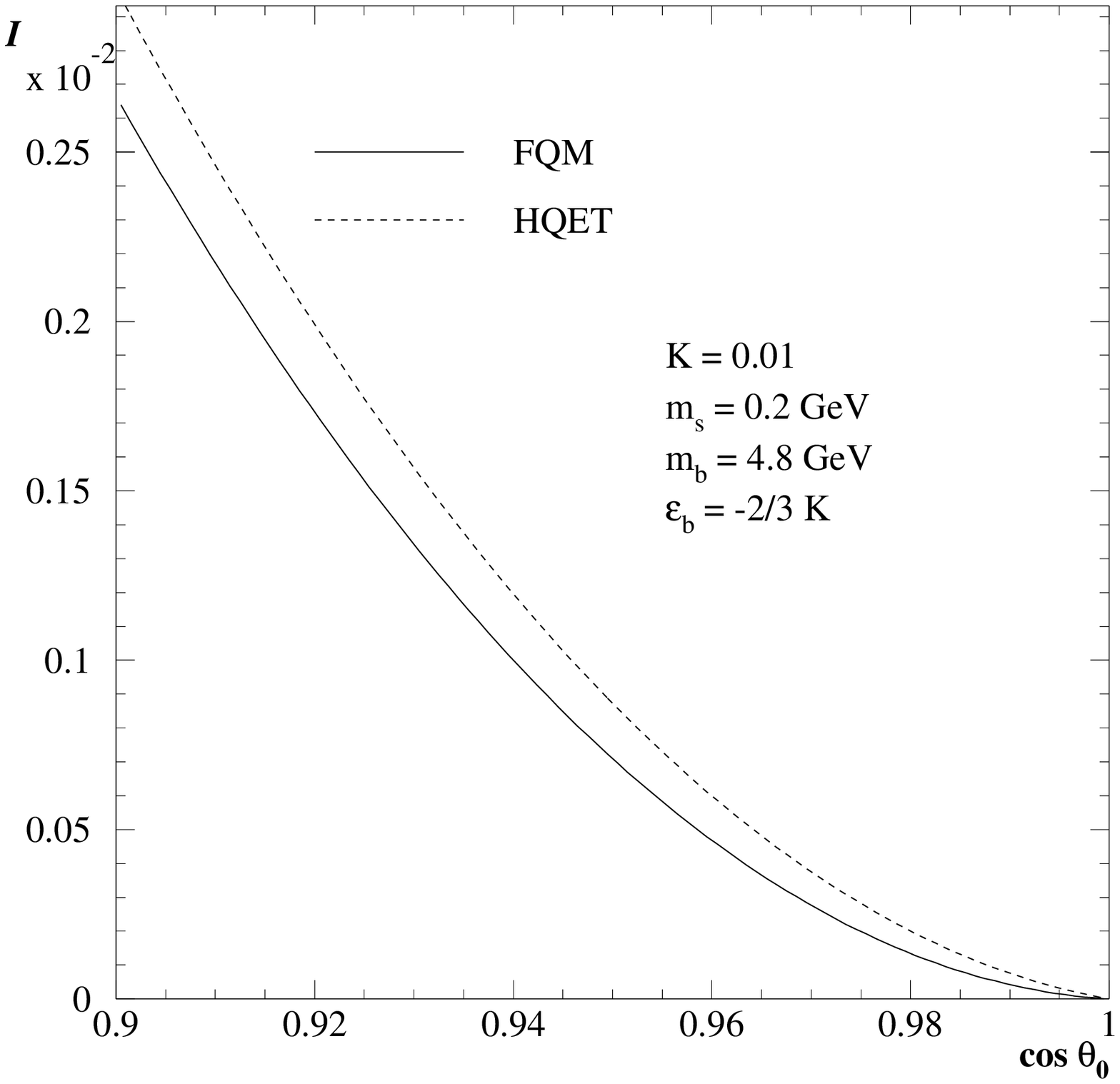}}
\caption{\label{fba}}
\end{figure}

{}From the angular distribution in \eq{decay}, we infer that the effect of
non-perturbative corrections is to reduce slightly the forward-backward
asymmetry in the photon emission. In \fig{fba}, we plot the fraction of decays
producing a photon in the forward cone $\cos\theta_0<\cos\theta<1$:
\be\label{asym}
I(\cos\theta_0)\equiv\frac{\int\limits_{\cos\theta_0}^1
d\cos\theta\,\displaystyle\frac{d\Gamma}{d\cos\theta}}{\int\limits_{-1}^{+1}
d\cos\theta\,\displaystyle\frac{d\Gamma}{d\cos\theta}}\ .
\ee
The result is compared with that in the free quark model.

{}From the results in Eqs.~(\ref{hel+}) and (\ref{hel-}), we obtain the
polarization of the photon as a function of its direction:
\be\label{pol}
P(\cos\theta) = \frac{d\Gamma_+/d\cos\theta - d\Gamma_-/d\cos\theta}
{d\Gamma_+/d\cos\theta + d\Gamma_-/d\cos\theta}\ .
\ee
Once again, as shown in \fig{pol}, corrections to the free quark model are
significant for photons emitted in the near-forward direction.

To summarize, the angular distribution of inclusively produced photons in the
decay $\LB$ of a polarized $\Lambda_b$ baryon can be calculated unambiguously
in HQET. This distribution tests the structure of the underlying \ha\ $\heff$
in a way that is not possible by studying the mesonic decay $B\to X_s\,\gamma$.
The angular distribution is a well defined observable which involves no
divergences of the type that appear in the calculation of the energy spectrum
(delta functions and derivatives thereof). Our calculation may be viewed as an
illustration of the utility (and limitation) of the HQET approach in
determining the effects of hadronic binding on the decay of a heavy quark.
The deviations from free quark decay in the reaction $\LB$ are found to be
globally small, but are significant for photons emitted in the forward
direction.
\bigskip

\centerline{\bf ACKNOWLEDGEMENTS}
\bigskip
We acknowledge helpful discussions with Gabi K\"opp, and
collaboration on a related paper \cite{gks}. This work has been supported
in part by the Deutsche Forschungsgemeinschaft (DFG) under
Grant No. Se 502/4--1, which we gratefully acknowledge.
%
%

\newpage
\centerline{\bf FIGURE CAPTIONS}
\begin{enumerate}
\item[\bf Figure 1] The fractional intensity $I$ of photons in the forward
cone $\cos\theta_0\le\cos\theta\le 1$.

\item[\bf Figure 2] The photon polarization $P$ in the inclusive $\Lambda_b$
decay as a function of the photon direction with $K= 0.01$ and
$\epsilon_b=-\frac{2}{3}K$.
\end{enumerate}
\newpage
%
%
%
%
\begin{figure}
\centerline{\epsfysize=12cm\epsffile{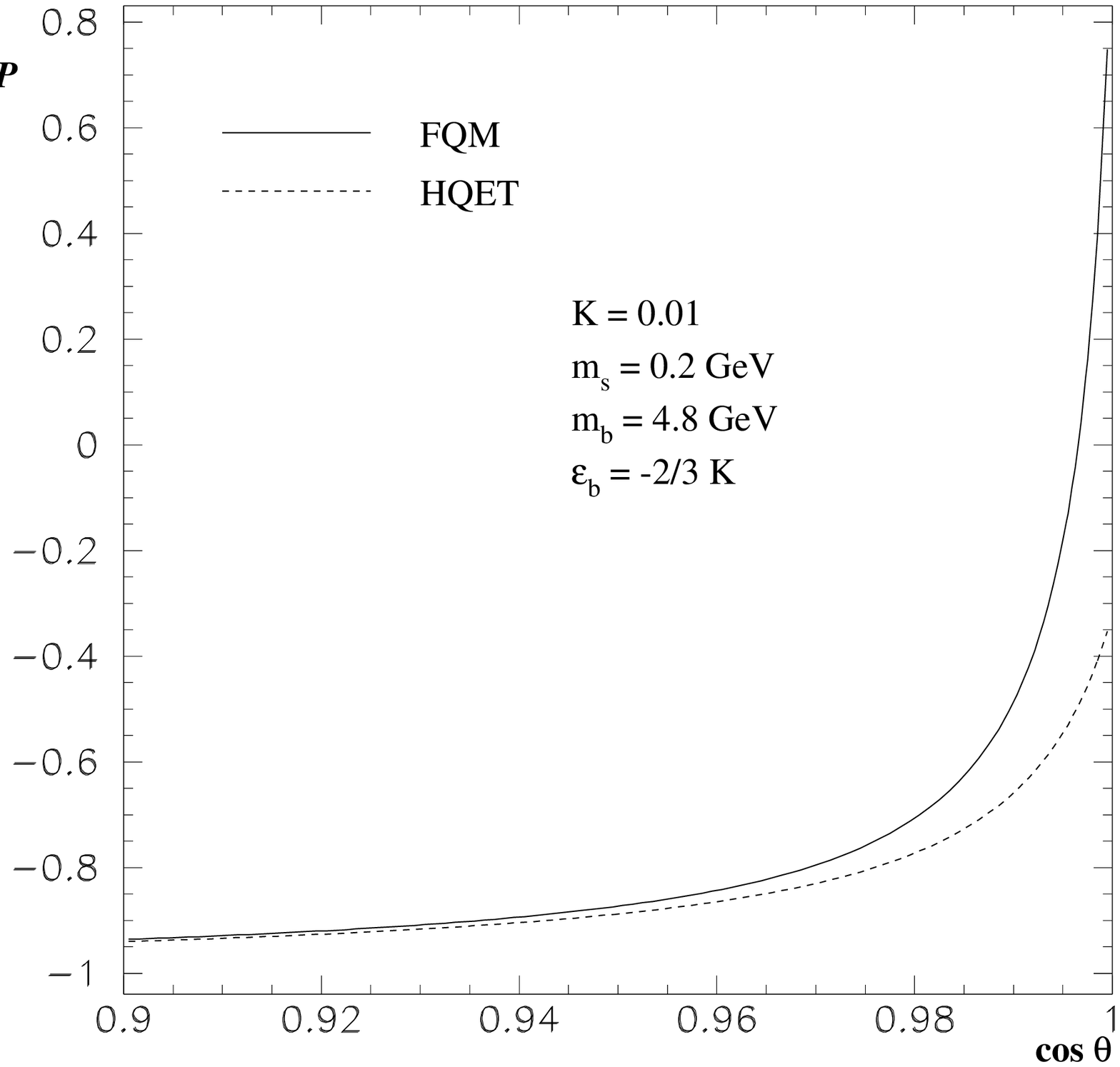}}
\caption{\label{pol}}
\end{figure}
\end{document}